# CITE THIS PAPER:

## 1.1.  Cite

Copy and paste a formatted citation or use one of the links to import into a bibliography manager.

# Hybrid Approach for Single Text Document Summarization using Statistical and Sentiment Features

## Abstract.

*Summarization is a way to represent same information in concise way with equal sense. This can be categorized in two type Abstractive and Extractive type. Our work is focused around Extractive summarization. A generic approach to extractive summarization is to consider sentence as an entity, score each sentence based on some indicative features to ascertain the quality of sentence for inclusion in summary. Sort the sentences on the score and consider top n sentences for summarization. Mostly statistical features have been used for scoring the sentences. We are proposing a hybrid model for a single text document summarization. This hybrid model is an extraction based approach, which is combination of Statistical and semantic technique. The hybrid model depends on the linear combination of statistical measures : sentence position, TF-IDF, Aggregate similarity, centroid, and semantic measure. Our idea to include sentiment analysis for salient sentence extraction is derived from the concept that emotion plays an important role in communication to effectively convey any message hence, it can play a vital role in text document summarization. For comparison we have generated five system summaries Proposed Work, MEAD system, Microsoft system, OPINOSIS system, and Human generated summary, and evaluation is done using ROUGE score.*

## 2. Introduction

Text document summarization playing an important role in IR (Information Retrieval) because, it condense a large pool of information into a concise form, through selecting the salient sentences and discards redundant sentences (or information) and we termed it as summarization process.

Radev et al. in [1] has defined a text summary as "a text that is produced from one or more texts that convey important information in the original texts, and that is no longer than half of the original text and usually significant less than that". As explained in [2] Automatic text document summarization is an interdisciplinary research area of computer science that includes AI (artificial intelligence), Data Mining, Statistics as well as Psychology. We can classify text doc summarization in two ways (by techniques) Abstractive summarization and Extractive summarization. Abstractive summarization is more human like a summary, which is the actual goal of Text document summarization. As defined by [3, 4] abstractive summarization needs three things as Information Fusion. Sentences Compression and Reformation. Abstractive summarization may contain new sentences, phrases, words even which are not present in the source document. Although till now a lot of research in happened in the last decades in the area of NLP (Natural language processing), NLG (Natural Language Generation), so much computing power increased, but still we are not near for abstractive summarization. The actual challenge is a generation of new sentences, new phases, along with produced summary must retain the same meaning as the same source document has. Extractive summarization based on extractive entities, entities may be sentence, sub part of sentence, phrase or a word. Our work is focused on extractive based technique.

In This paper we are proposing a hybrid method for single text document summarization, which is linear combination of statistical features as used in [5, 8, 7, 6] and new kind of semantic feature i.e. sentiment analysis. The idea which is used in this paper has been derived from different papers like for statistical features and their collective sum obtained from [5, 8], centroid measure are taken from [7,6]. To include sentiment analysis is derived from the concept that emotion plays an important role in communication to effectively convey any message hence, it can play a vital role in text document summarization.

Outline of paper looks like, in section 2 we are presenting categorized literature work done in recent years, section 3 contains features used for summarization purpose, section 4 contain summarization algorithm and detail approach, in section 5 we are presenting corpus description with statistical and linguistic statistic, section 6 showing some experiments and results, in section 7 is about conclusion.

## 3.    Related Work

According to M. Ramiz summarization in [9] is defined as a three steps process (1) Analysis of text. (2) Transformation- as summary representation, and (3) Synthesis- produce an appropriate summary. Eduard Hovy and Chin-Yew Lin [10] introduced SUMMARIST system to create a robust text summarization system, system that works on three phases which can describe in form of an equation like "Summarization = Topic Identification + Interpretation + Generation".

A lot of research done in the direction of Extraction based approaches. In extractive summarization the important the task is to find informative sentences, a subpart of sentence or phrase and include these extractive elements into the summary. Here we are presenting work done in two categories (1) early work done and, (2) work done in recent years. In our views these are three works done initially, that provides direction of Text Document Summarization (Extractive), explained below

The early work in Document Summarization started on "Single Document Summarization", by H. P. Luhn [11], he proposed a frequency based model, in which frequency of word play crucial role, to decide importance of any sentence in the given document. Another work of P. Baxendale [12], had been introduced a statistical model based on sentence position. In his research, he found that, starting and ending sentences became salient sentences for summarization, but this is not better for every document like scientific research paper but good for newspapers summarization. H.P. Edmondson [13] also proposed an effective technique for document summarization. At first, Edmonson designed some rules for manual extraction, then rules were applied to 400 technical documents. Edmondson considers four features (1) sentences position, (2) frequency of word, (3) presence of cue words, and (4) the skeleton of the document.   The work was done almost manual. After these early work lot of work done in this discipline some are available in [1, 14, 15], here in the next section we are presenting only work done in recent years.

Query focused summarization is a special case of document summarization, in which summary purposely demands, to be biased according to the user query. You Ouyang et.al [16] used SVR (Support Vector Regression) to calculate the importance of the sentences in a given document. Another query focused summarization, multi document summarization done by Carbonell, J., & Goldstein in [17].

Researcher done a lot of work in the multi document summarization field and many more like in [18] considering this is as a global/multi optimization problem which requires simultaneous optimization of more than one objective function. Rasim M. Alguliev [18], in which the objective function is a weighted combination of (1) content coverage, and (2) for redundancy objectives. In another work [19] they proposed CDDS based summarization with two objectives diversity and coverage. In summarization, similarity evaluation among of sentences is a laborious task because of (1) complex sentence structure, and (2) lack of extra information so, Ming Che Lee [20] had been proposed "Transformed Vector Space" model based on Word-Net. Due to improper ordering of information there is possibility that it can confuse the readers as well as can degrade the readability of the summary. To maintain the association and order of sentences, Danushka Bollegala et.al [21] defined four criteria (1) chronology, (2) Topical-closeness (3) Precedence (4) Succession. These all four criteria are combined into a single criteria by using a supervised learning approach.

Redundancy can be defined as a multiplicity of sentences, sub sentences or information. Coverage and redundancy are reciprocal to each other. Summarization objective is maximum coverage and minimum redundancy. Kamal Sarkar [22] gave a simple approach to include a sentence in summary one by one based on modified cosine similarity threshold. To include sentences in the summary, system first selects the most top rank sentence, include it in the summary and this process is repeated for remaining sentences. Next sentence is included in the summary if similarity between sentences and summary is less than some threshold, otherwise the sentence is not included in the summary and algorithm stops when required summary length is reached. We are using this model in our implementation to handle redundancy. Another model MMR [23] is popularly used (especially in with a given query) to reduce redundancy. The MMR (Maximal Marginal Relevance) criteria, "strives to reduce redundancy while maintaining query relevance in re-ranking retrieved documents and in selecting appropriate passages for text summarization". This technique gives better result for Multi Document Summarization. Rasim M. Alguliev et.al [24] proposed an unsupervised text summarization model which can be used for Single as well as Multi Document Summarization. In simple terms, the problem is treated as a Multi objective problem, where the objective is to optimize three objectives (1) Relevance, (2) Redundancy and (3) Length.

To find a good summary lot of work done, but to decide the quality of the summary still a challenging task. Research is done by Goldstein in [23] he conclusion that (1) "even human judgment of the quality of a summary varies from person to person", (2) only little overlap among the sentences picked by people, (3) "human judgment usually doesn't find concurrence on the quality of a given summary". Hence it is sometimes difficult to judge the quality of the summary. For evaluation most researcher use the "Recall Oriented Understudy for Gisting Evaluation" (ROUGE) introduced by Lin [25] and this has been officially adopted by DUC for summarizer evaluation. ROUGE compares system generated summary with different model summaries. It has been considered that ROUGE is an effective approach to measure document summarizes so widely accept. ROUGE measures, overlap words between the system summary and standard summary (gold summary/human summary). Overlapping words are measured based on N-gram co-occurrence statistics, where N-gram can be defined as the continuous sequence of N words. Multiple ROUGE metrics has been defined for different value of N and different models (like LCS, weighted). Standard ROUGE-N is defined by:

$$ROUGE - N = \frac{\sum_{S\epsilon\{ReferencesSummaries\}}\sum_{gram_n \epsilon S} Count_{match}(N-gram)}{\sum_{S\epsilon\{ReferencesSummaries\}}\sum_{gram_n \epsilon S} Count(N-gram)} \qquad (1)$$

Here N stands for the length of the N-gram, Count(N-gram) is the number of N-grams present in the reference summaries, and the maximum number of N-grams co-occurring in the system summary, the set of reference summaries is Countmatch (N-gram) ROUGE measures generally gives three basic score Precision, Recall, and F-Score. Since ROUGE-1 score is not a sufficient indicator of summarizer performance, so another variation of ROUGE is; ROUGE-N, ROUGE-L, ROUGE-W, ROUGE S*, ROUGE SU*. In our evaluation we are using fourteen ROUGE measure (N= 1to 10, L, W, S*, and SU*). For other variation kindly follow [25].

## 4.    Proposed Model for text document summarization

In this section we are presenting Features used in our sentence selection approach in section 3.1 , and detailed approach in section 3.2.

### 4.1.    Features

We are proposing a hybrid model for salient sentence extraction for "Single Text Document Summarization" based on two types of features statistical based features i.e. Location, Frequency (TF-IDF), Aggregate Similarity, Centroid and Semantic based feature (sentiment ).

**Statistical Features used**

*A.) The location Feature (Score1).*

P. Baxendale in [12], introduced a feature based on "Sentence Position". Although his work was almost manual but, later on this measure used widely in sentence scoring, he proposed that leading sentences of an article are important. Model which we are using given below, where N is total number of sentences. The used model is: (Where: 1<i<N, and $Score(S_i)$ =(0,1] )

$$Score(S_i) = 1 - \frac{i-1}{N} \qquad (2)$$

*B.) The aggregation similarity Feature (score2).*

Kim et al.[26] defined aggregate similarity as, "the score of a sentence is as the sum of similarities with other all sentence vectors in document vector space model". It is given by

$$Sim(S_i, S_j) = \sum_{k=1}^{n} W_{ik}.W_{jk} \qquad (3)$$

$$\text{Score}(\text{S}_i) = \sum_{j=1, j \neq i}^{n} \left( \text{Sim} \left( \text{S}_i , \text{S}_j \right) \right) \qquad (4)$$

Where $W_{ik}$ is defined as the binary weight ok kth word in $i^{th}$ sentence. Similarity measure plays an important role in text document summarization, even studied different similarity measure affects the outcome. In our implementation, using Cosine Similarity that is widely used. The cosine measure between two sentences $S_i = [W_{i1}, W_{i2},..W_{im}]$ and $S_j = [ W_{j1}, W_{j2},....W_{jm}]$. Standard Cosine similarity measure gives by following formula which is used in our implementation is below.

$$Sim(S_i, S_j) = \frac{\sum_{k=1}^{m} W_{ik} . W_{jk}}{\sqrt{\sum_{k=1}^{m} W_{ik}^2 . \sum_{k=1}^{m} W_{jk}^2}} \quad \text{i,j=1 to n} \qquad (5)$$

### C.) Frequency Feature (score3).

The early work in Document Summarization started on "Single Document Summarization", by H. P. Luhn [11] at IBM in the 1950s. Luhn proposed a frequency based model, frequency of word play a crucial role, to decide the importance of any word or sentence in a given document. In our method, we are using the traditional method of "TF-IDF" measure is defined as below, i.e. TF stands for term frequency, IDF for inverse document frequency.

$$W_i = (TF_i) \times (IDF_i) = tf_i \times \log \frac{ND}{df_i} \qquad (6)$$

Where, $TF_i$ is the term frequency of $i_{th}$ word in the document, ND represents total number of documents, and $IDF_i$ is the document frequency of $i^{th}$ word in the whole data set. In our implementation to calculate importance of word $W_i$, for TF we considering the sentence as a document and for IDF entire document as a Data set.

### D.) Centroid Feature (score4).

Radev et.al [7] defined centroid as, "a centroid is a set of words that are statistically important to a cluster of documents". As such, centroids can be used both to identify salient sentences in a cluster and classify relevant documents. The centroid score $C_i$ for sentence $S_i$ is computed, as the sum of the centroid scores $C_{w,i}$ of all words appeared in the particular sentence.

$$C_i(S_i) = \sum_w C_{w,i} \qquad (7)$$

## Sentiment Feature or Semantic Feature used

### E.) Sentiment Feature (score5).

In previous sections, we mentioned statistical measures used by us. We are calling this feature as a semantic feature because in this a set of things are related to one other. Semantic Summary generation may be done using shallow level analysis and deep level analysis as defined by [27]. In shallow approach to most analysis done on sentence level is syntactic, but important to note that, word level analysis may be semantic level and, in deep analysis at least a sentential semantic level of representation is done. So our approach i.e. sentiment feature is semantic and low level analysis (because at the entity level).

For finding sentiment score for a sentence, fist we find all entities present in a sentence then find sentiment score of each entity and then do sum of all entity's sentiment score (i.e. sentiment strength). If sentiment of entity is neutral then we scoring it as 0, if entity's sentiment is positive then considering as same and adding to find the total score of a sentence but if sentiment score is negative we multiplying it by -1 to covert in positive score then adding this score to find total score. Reason for considering negative score to positive score is that we are interested only in sentiment strength which may be positive or negative i.e. if sentiment score of an entity if "-.523" it means sentiment of entity is negative and strength is ".523". Detail procedure are explained in section 5.1's Fifth feature. Here | A | representing mode (A) i.e. |-A | = | A | = A.

$$\text{Score5} = \sum_{i=1}^{n} | Sentiment(Entity_i)| \qquad (8)$$

# 5.    Summarization Procedure

Our summarization approach is based on salient sentence extraction. The importance of any sentence is decided by the combined score given by the sum of Statistical Measures and Semantic measure. In next step we are explaining our approach (algorithm) used in this paper. Basically work of summarization can be dived in three PASS (1) Sentence Scoring, (2) Sentence extraction and (3) Evaluation.

## 5.1.    Algorithm

**PASS1:** Sentence scoring according linear combinations of different measures.
**PASS2:** Salient sentence Extraction (Summary Generation).
**PASS 3:** Evaluation of Summary.

**PASS 1:** Sentence Scoring
**Input:** Documents
**Output:** Scored sentences
Step 1:    Score the sentence given with 5 different measures. Outcome is M × N Matrix (M no. of sentences, N no. of measures) (1) Aggregate cosine similarity, (2) Position, (3) Sentiment of sentence, (4) Centroid score of sentence, (5) TF×IDF.
Step 2:    Normalized columns of matrix
Step 3:    Add all the features for every sentence, we calling this sum as a score of the sentence.
Step 4:    Sort according to score, a highest score representing most significant sentence.

**PASS 2:** Algorithm for Redundancy
**Input:** Number of sentences descending according to total score
**Output:** Extracted sentences
**Parameter Initialization:** (1) Summary= "" // Empty summary, (2) Similarity Threshold "θ", (3) L // required length summary.
Step 1:    Summary = (Topmost scored sentence)
Step 2:    For i=1 to (number of sentences)
if (Similarity(Summary, i$^{th}$ sentence) <θ ) AND (Length (summary) < L) Then
Summary= Summary + i$^{th}$ sentence
Step 3:    Rearrange Summary sentences, as given in Source     Document for maintain cohesiveness.

**PASS 3:**  Evaluation of Summary
**Input:** Different summaries as "Standard Summaries" and "Peer summaries"
**Output:** Precision, Recall and F-score
Step 1:    Generate different summary, different length using MEAD, MICROSOFT, OPINOSIS, HUMAN (5 human) and Our-Proposed algorithm.
Step2:    For experiment 1
Model summary: MEAD, MICROSOFT, OPINOSIS
Peer summary: Our-Proposed algorithm
For Experiment 2:
Model summary: Human generated summary
Peer summary: MEAD, MICROSOFT, OPINOSIS, Our-Proposed method
Step 3:    Used Rouge-N (N=1 to 10), ROUGE-L, ROUGE-W (we set W=1.5), ROUGE S*, ROUGE SU* measure to find Precision, Recall, and F-Measure.

## 5.2. Detailed Approach Description

Here we are describing detail approach used as the procedure described above.

**PASS 1.** Sentence scoring and Extraction:

In Algorithm defined in the previous section most things are covered and gives the main idea of Algorithm, still some micro points are needs to specify. Pass 1 is the sum of the linear combination of five different measures, four are statistical dependent (i.e. Aggregate Similarity, Position, TF ×IDF, Centroid) and fifth measure is semantic dependent (i.e. Sentiment).

**(1) First feature,** is Position of sentences, Position is an important indicator for important sentence and it has been analyzed that first or leading sentences mostly contains important information. The score1=position=1-(((i-1))/ (N)). Position score for some sentence index 0 to 55 are given in Table-1's second column.

**(2) Second feature,** TF × IDF approach we are using the standard formula as defined in the previous section. Normalized TF-IDF score are given in Table-1's third columns.

**(3) Third feature,** is an Aggregate similarity (cosine) score of a sentence vector can be calculated as the, "sum of similarities with other all sentence vectors in document Vector Space Model". The significance of this is to find a sentences which are highly similar to all other sentences. After representing all sentences in a vector space, and then find vector cosine similarity with all other sentences as defined standard formula in the section. Normalized Aggregate cosine similarity in Table-1 column four.

Since other scores (Centroid, Position, Sentiment) are between [0,1] so we need to normalized score. Normalization of values means "adjusting to values measured on different scales to one notionally common scale"[28] that removes chance to be bias w.r.t. some values. in our implementation we are just using column normalization instead of matrix normalization. **Normalization of a column** vector X=[$X_1$, $X_2$.......$X_n$] is done using equation 9. Where $X_i$ is the $i^{th}$ element in the column, and n is the size of the column.

$$X_i = X_i \times \frac{1}{\sqrt{\sum_{i=1}^{n} X_i^2}} \qquad (9)$$

$$A = \begin{bmatrix} 1 & 1 & .4 \\ 2 & 4 & .4 \\ 3 & 5 & .3 \end{bmatrix} \quad B = \begin{bmatrix} 1 \times \frac{1}{\sqrt{1^2+2^2+3^2}} & 1 \times \frac{1}{\sqrt{1^2+4^2+4^2}} & .4 \\ 2 \times \frac{1}{\sqrt{1^2+2^2+3^2}} & 4 \times \frac{1}{\sqrt{1^2+4^2+5^2}} & .4 \\ 5 \times \frac{1}{\sqrt{1^2+2^2+3^2}} & 5 \times \frac{1}{\sqrt{1^2+4+5^2}} & .3 \end{bmatrix}$$

Let A is a given matrix, which size is 3×3 and column one and two has doesn't have values between [0,1] , then we are doing normalization of only column one and two but not third column and B is the give normalized matrix in our case.

**(4) Fourth feature,** is centroid based, D.R.Radev [7] defined as "Centroid as a set of words that are statistical important to a cluster of documents". In our approach using MEAD centroid score output as our input. The centroid value of a sentence, is given by summation of each word's centroid score present in the sentence.

**(5) Fifth feature,** is "Sentiment score", this is a novelty in our work, to find this feature we depends on Alchemy API which is available at http://www.alchemyapi.com/. Our consideration is that it is finding sentiment score is a semantic approach and fall under shallow level approach as defined in section 3.2. For any sentence or words we can define three kind of sentiment (a) Neutral, (b) Negative, and (c) Positive. Neutral sentiment value mean that words or that sentence sentiment score is zero, most important to note that it is easy to find sentiment score based on cue word like good, bad, pleasant etc., but still due to so much complexity in text, words, limitation of NLP etc. it is not possible to find correct sentiment score, sometime even it is also not possible to detect sentiment due to hidden sentiments,

**Document 1:** "NAAC Accredited JNU with the CGPA of 3.91 on four point scale of A grade (highest grade by NAAC)", and Sentiment of this is "NEUTRAL".

**Document 2:** "JNU ranked in top 100 in Times Higher Education Asia and BRICS Top Ranking", and Sentiment of this document is Positive and score is 0.499033. **[NOTE-2 NAAC stands for National Assessment and Accreditation Council (NAAC), and BRICS stands for five nations Brazil, Russia, India, China, and South Africa]**

Here Document 1 and Document 2 both representing positive news about JNU. But the sentiment of document 1 is neutral and sentiment of document 2 is Positive with .499033 score. We still have to discover approach which can find correct sentiment (hidden sentiment). Some results are displayed in Table 1 with sentiment results.

In our implementation to find sentiment score of a sentence we are using alchemy API, first finding all entities present in the sentence and their sentiment score, then we add all entity's sentiment score. For example consider document number 49 "Meanwhile, BJP spokesperson Prakash Javadekar has said that party president...etc " has five entities as follow (1) Prakash Javadekar : Person Name: -0.212091 (2) Rajnath Singh : Person Name: -0.212091 (3) Uttarakhand: State/County: -0.212091 (4) BJP: Company : -0.212091, (5) president: JobTitle: -0.212091 [NOTE-3 Triplet X: Y: Z representing, X is Entity Name, Y is Entity Type, Z Sentiment Score.], and to give sentiment score of sentence 49 we add all, Score5= $\sum_{i=1}^{n} Sentiment(Entity_i)$ so sentiment score (score5)= -1.060455 for sentence 49. But, if we see the Table 1 result in row 49 and column 3 (sentiment score) the value is 1.060455. Here we are considering only positive sentiment scores, if entity sentiment score is negative then by multiplying -1, to convert it into positive score. The obvious goal of this procedure is to give equal importance if magnitude is same. Let consider one childhood story **"The fox and the grapes", in figure 1.**

# Figure 1 "The Fox and The Graps story"

One afternoon a fox was walking through the forest and spotted a bunch of grapes hanging from over a lofty branch. Just these sweat and juicy grapes to quench my thirst, he thought. Taking a few steps back, the fox jumped and just missed the hanging grapes. Again the fox took a few paces back and tried to reach them but still failed. Finally, giving up, the fox turned up his nose and said, they're probably sour anyway, and proceeded to walk away".

If here we consider two sentences (given below , document 3 and 4) as document, both are important in the story and about same things grapes.

**Document 3:** Just these sweat and juicy grapes to quench my thirst, and

**Document 4:** They're probably sour anyway.

With Alchemy system if we find a sentiment of both these sentences, then the sentiment of document 3 is positive and the score is 0.707112 where, the sentiment of document 4 is negative with -0.598391. For us only magnitude is important, reasons to consider as + value are (1)we are interested to find sentiment strength, it may be negative or positive and both are important for us, and (2) if we will add negative score to find total score then value will reduce.

In next step we are finding the total score of a sentence, by adding all scores. The total score can be represented by equation 10, given below. In our implementation $w_k = 1 \ (k = 1 \ to \ n)$. Detail result of individual score is given in Table 1, and last column is total score of all scores.

$$TOTALSCORE(S_i) = \sum_{k=1}^{5} w_k \times score_k. \qquad (10)$$

*Table 1: Different Features scores and Total Score for sentences*

| Sent No. | Position Score | TF * IDF | Aggregate Cosine sim. | Centroid Score | sentiment Score | SUM OF ALL |
|---|---|---|---|---|---|---|
| 0 | 1 | 0.175146269 | 0.154823564 | 0.750856924 | 0.661311 | 2.742137758 |
| 1 | 0.98245614 | 0.127104001 | 0.156975892 | 0.307405503 | 0 | 1.573941537 |
| 2 | 0.964912281 | 0.170528144 | 0.185519412 | 0.50702023 | 0.394824 | 2.222804067 |
| 3 | 0.947368421 | 0.182701719 | 0.174777773 | 0.663083509 | 0.217356 | 2.185287422 |
| 4 | 0.929824561 | 0.119835913 | 0.144151017 | 0.440031566 | 0 | 1.633843058 |
| 5 | 0.912280702 | 0.184106205 | 0.176538454 | 1 | 0.389024 | 2.661949361 |
| 6 | 0.894736842 | 0.128152365 | 0.144782619 | 0.353263066 | 0 | 1.520934892 |
| 7 | 0.877192982 | 0.172677585 | 0.14243688 | 0.423917828 | 0.889318 | 2.505543275 |
| 27 | 0.526315789 | 0.098431452 | 0.034336443 | 0.120511453 | 0 | 0.779595137 |
| 49 | 0.140350877 | 0.148292229 | 0.122318812 | 0.423393814 | 1.060455 | 1.894810731 |
| 50 | 0.122807018 | 0.091023788 | 0.054947802 | 0.048475042 | 0 | 0.31725365 |
| 51 | 0.105263158 | 0.076719171 | 0.044899933 | 0.102930791 | 0 | 0.329813053 |
| 52 | 0.087719298 | 0.126084523 | 0.149793298 | 0.206132535 | 0.327554 | 0.897283655 |
| 53 | 0.070175439 | 0.127388217 | 0.037996007 | 0.137435141 | 0.363308 | 0.736302803 |
| 54 | 0.052631579 | 0.113474586 | 0.064967719 | 0.114433381 | 0 | 0.345507265 |
| 55 | 0.035087719 | 0.100535505 | 0.066218503 | 0.247865211 | 0 | 0.449706939 |

**PASS 2. Redundancy:**

To remove redundancy we used same model as proposed by Sarkar in [22] which, the top most sentence (according to total score defined in equation) is add in summary, we add next sentence in the summary if similarity is less than threshold θ. Algorithm is described in section 5 pass-2, input in this pass is a number of sentences which are sorted according to descending total score. We need to initialize some parameter to get the desired summary, parameter like summary initially empty, given similarity threshold θ and L for desired length summary. Even in our system L means maximum length of desired summary, but due to limitation (here length of sentences), we can guaranteed minimum (maximum (length (summary))). We will add the next sentence in the summary, if still summary length is less than L and similarity (new-sentence, summary) <θ. The Output of this step is a summary with minimal redundancy and length nearly equal to L but the position of the sentence is zigzag that lost the sequence and cohesiveness. To maintain the sequence we need to rearrange the sentences according to given in initial index.

In Table 2 we are representing the summary generated by our system, in which similarity threshold θ is .1, and desired summary length is 15 %. We can define arbitrary L in a number of words or percentage of summary required. Here we chosen θ small. if we put θ large like .4 or .5 then the sentences, which are in coming in summary, will depend only on  the total score (as in Table 1).  In other words the summary is only depended on totalscores as shown in Table 1, but our objective is also to minimize redundancy. [NOTE 4: Before calculating sim(new sentence, summary), we are eliminating stopwords, stopwords play a big role to increase the similarity between two sentences. with different stopwords list we will get different similarity score].  MEAD, Microsoft, and Our-Model generated summary with different length are shown below in Table 2,3,4. The truth is we don't have the detail of Microsoft generated a summary. This summarizer is inbuilt inside Microsoft office package. When we observed then find little unhappy that, Microsoft summarizer is not reducing redundancy sentence 4 and 18 are almost similar sentences, see Table 3.

**PASS 3. Evaluation:**

Goldstein in [23] he concluded two things (1) "even human judgment of the quality of a summary varies from person to person", (2) "human judgment usually doesn't find concurrence on the quality of a given summary", hence it is difficult to judge the quality of a summary. For evaluation of any summary we need two summary first one is a system generated summary and other summary is user generated (Model summary or Standard Summary).

To generate different model summary we used three approaches (1) we given our text data set to 5 person and tell them to write a summary in about to 20% to 40% words, (2) We generate summary by MEAD tool, in this approach we apply linear combination of position and centroid score, score(mead)=(w1×centroid)+(w2×position)  with w1=w2=1, and (3) third model summary is generated by OPINOSIS[29] (summary given in figure 3) (4) Microsoft System.

To evaluate summary we are using ROUGE evaluation package. ROUGE is adopted by DUC for official evaluation metric for both single text document summarization and multi document summarization. ROUGE finds Recall, Precision, and F-Score for evaluation results. Based on N-gram co-occurrence statistic it (ROUGE) measures how much the system generated summary (machine summary) overlaps with the standard summary (human summaries/model summary). Where an N-gram is a contiguous sequence of N words. In our evaluation we are adopting different measures of ROUGE, as ROUGE-N (N=1 to 10), ROUGE-W, ROUGE-L, ROUGE-S*, and ROUGE-SU*.

## 6.    Corpus Description

"In 16 June 2013, was a multi-day cloudburst centered on the North Indian state of Uttarakhand caused devastating floods along with landslides and became the country's worst Natural Disaster. Though some parts of Western Nepal, Tibet, Himachal Pradesh, Haryana, Delhi and Uttar-Pradesh in India experienced the flood, over 95% of the casualties occurred only in Uttarakhand. As of 16 July 2013, according to figures provided by the Uttarakhand Government, more than 5,700 people were presumed dead"[30]. Corpus is self-designed, taken from various newspapers ex. "The Hindu", "Times of India". This Dataset is also published in paper C.S.Yadav [31,32]. Here we are showing some statistically and linguistic statics about our Data Set used.

### 6.1.    Statistical statistics

Total No. of Sentences in document 56,
Length of document after stop word removed: 1007,
Total number of distinct words: 506,
Minimum sentence Length 6 words,
Maximum sentence Length 57 words,
Average sentence length is 1454/56 = 25.96.
In our experiment we used SQL stopword list which is available at [33].By seeing the figure 2, we can interpret that 45 sentences are between length 10 and 40, and 36 sentences are between length 15 and 35.

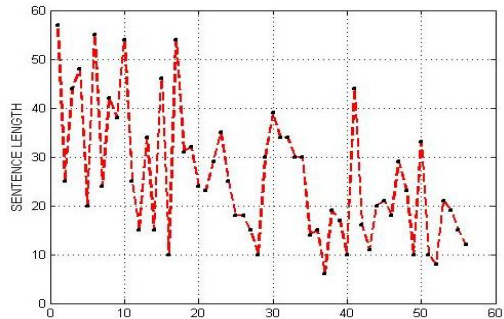

Figure 2: Sentence Length (Y-axis) Vs Sentence Number (X-axis)

### 6.2.    Linguistic statistics

(1) 'NN': 208, 'NNP': 196, 'NNS': 131 ;
(2) 'DT': 150 ; (3) 'JJ': 70, 'JJR': 7 'JJS': 2 ;
(4)  'VB': 37, 'VBN': 64,  'VBD': 48 'VBZ': 38, ',': 38, 'VBG': 37, , 'VBP': 22
where different abbreviation stands for [ "NN-Noun, singular/mass, NNS-Nounplural,  NNP-Proper noun singular, NNPS-Proper noun plural,  VB-Verb, VBD-verb past tense, VBG-verb gerund, VBN-verb past participle, VBP-verb non-3rd person singular, VBZ-verb 3rd person singular, JJ-Adjective, JJR-Adjective comparative, JJS-Adjective superlative, DT-Determinant"]. [**NOTE 1**: 'X':10 means, X is entity type and 10 is its count].

# 7.    Experiment and Results

In this section we are presenting two experiment done on mentioned dataset.

## 7.1.    Experiment 1

As explained in section 5's pass 3 we created 4 types of model summary (1) Human summary (via we gave data set to 5 persons to summarize it, based on their experience with instruction to summarize it within 20% to 40% words length. Due to limitations and user experiences, the generated summary varies from 24 % to 52% words length), (2) MEAD, (3) Microsoft, and (4) OPINOSIS system. Different system generated summary are given in Table 2,3,4. Since OPINOSIS summarizer is abstractive type, in figure 3 we are giving summarization result length 10% summary generated by OPINOSIS System. Table 2, 3 ,4 presenting different summaries generated by different systems.

*Table 2: Our-System generated summary (using proposed Algorithm).*

| Sent No | Extracted sentences (Our proposed Model) 15% summary length |
|---|---|
| 0 | "Uttarakhand-Flood Missing untraced till July 15 will be presumed dead: Bahuguna acing giants time, the Uttarakhand government on Thursday decided that those missing in the flood-ravaged state will be presumed dead if they remain untraced till July 15 and asked officials to remain vigilant in the wake of warning of heavy rains over the next two days". |
| 5 | "A team of seven mountaineers is also engaged in a combing operation in areas adjoining the shrine in search of bodies while over  50 members of a team |
| 27 | Fifty-five helicopters have been pressed into service for rescue work". |
| 35 | "There has been large scale destruction of property and loss of lives in this disaster". |
| 37 | "The CRPF rank and file joins the countrymen in conveying its deepest concern for the victims of the tragedy". |
| 42 | "We are keeping them in the morgue and documenting their details". |
| 44 | "Central Army Commander Lt Gen Anil Chait said on Friday that about 8,000 to 9,000 people are still stranded in Badrinath". |
| 52 | "He also praised the efforts of the armed forces and the Indo-Tibetan Border Police, saying they were doing a laudable job". |

*Table 3: Microsoft system generated summary.*

| SENT NO | Extracted sentences (microsoft summary) 10 % |
|---|---|
| 4 | "Meanwhile, the Indian Air Force flew 70 civil administration personnel to the Kedarnath temple premises to clean the surroundings there |
| 18 | "The Indian Air Force (IAF) today flew 70 civil administration personnel to the Kedarnath temple premises to clean the surroundings after |
| 26 | "The Indian Air Force (IAF) has deployed 13 more aircraft for relief and rescue operations". |
| 29 | "New Delhi: Indian Air Force has airlifted over 18,000 persons and dropped more than 3 lakh kg of relief material in flood-hit Uttarakhand |
| 32 | "The Central Reserve Police Force (CRPF) on Saturday announced it will contribute one day's salary of its personnel to the Prime Minist |
| 41 | "Rescue teams and police personnel have recovered 48 dead bodies from the River Ganga in Haridwar". |

*Table 4: MEAD system generated summary.*

| SENT NO | Extracted sentences (mead system summary) 15 % |
|---|---|
| 0 | Uttarakhand-Flood Missing untraced till July 15 will be presumed dead: Bahuguna acing gainst time, the Uttarakhand government on Thursday decided that those missing in th |
| 1 | "Chief Minister Vijay Bahuguna said the exact number of people missing after the tragedy is 3,064 and the deadline for finding them is July 15". |
| 2 | "Considering the magnitude of the crisis, the state Cabinet has decided that if the missing persons are not found by July 15, we will presume that they are dead and the rocess of paying compensation to their next of kin will begin," he said. |
| 3 | "With the MeT department issuing a warning of heavy rains at places in Kumaon region over the next two days, Bahuguna said that for the next 50 hours the administration needs to be highly vigilant, adding 250 National Disaster Response Force personnel have been deployed in these areas". |
| 4 | "Meanwhile, the Indian Air Force flew 70 civil administration personnel to the Kedarnath temple premises to clean the surroundings there". |
| 5 | "A team of seven mountaineers is also engaged in a combing operation in areas adjoining the shrine in search of bodies while over 50 members of a team of experts and volunteers is stationed in Kedarnath to clean the temple premises of tonnes of debris under which more bodies may be lying, an official said". |
| 8 | "Mass cremation of bodies in Kedarghati held up for the past few days on Thursday started with 23 more consigned to flames, taking the number of bodies disposed of so far to 59 even as a team of experts worked on removal of debris and extricating bodies from under them at the Himalayan shrine". |
| 14 | "Despite continuing bad weather in affected areas amid a MeT department prediction of heavy rains in the next 48 hours at places, efforts were on to airdrop relief material in affected villages totally cut off after the calamity in the worst-hit Rudraprayag, Chamoli and Uttarkashi districts". |
| 29 | "New Delhi: Indian Air Force has airlifted over 18,000 persons and dropped more than 3 lakh kg of relief material in flood-hit Uttarakhand since June 17 in its biggest ever helico |

# Figure 3 OPINOSIS generated summary

> *Indian air force has deployed 13 more aircraft for relief and rescue operations. a total of 93 sorties and dropped about 12,000 kgs of relief and equipment , said .Indian air force flew 70 civil administration personnel to the temple premises to the surroundings there . for the pastfew days on Thursday started with 23 more consigned to flames .people have so far been evacuated from the flood and landslide-hit areas of Uttarakhand so far and so far .gen chait said on Friday that about 8,000 to 9,000 people are still stranded in Badrinath .efforts to help those in distress in different inaccessible parts of the state and the hill state .rs18 crore to support the victims of the massive.*

In the first experiment we took our own summary (generated by algorithm discussed in section 4) as system generate summary and another summary as Model summary. In next step we find different Rouge scores (N=1 to 10, ROUGE-L, ROUGE-W where W=1.5, ROUGE S* and ROUGE -SU*) as defined by [25]. ROUGE scores is given by formula (1) defined in section 2. It measures 3 things Recall, Precision and F-Score for any System generated summary and Model Summary (or Reference summary). Recall and Precision are given in a slightly different way as defined by [34].

$$Precision = \frac{Count_{matc\ h}(Sentence\ )}{Count_{candidate}\ (Sentence\ )} \qquad (11)$$

$$Recall = \frac{Count_{matc\ h}(Sentence\ )}{Count_{bestsentence}\ (Sentence\ )} \qquad (12)$$

We are comparing our system generated summary, with other's (as model summary) same length summary. Result of this Experiment-1 is given is Table 5, Table 6, and Table 7, for 10%, 20%, 30% length respectively (due to limitation of space we providing only three tables).  Figure 6 showing F Measure with different model summaries of length of nearly 30% and our summary length is nearly 27%. In simple term we can define "High precision means that an algorithm retrieved substantially more relevant than irrelevant" and, "High recall means that an algorithm return most of the relevant result"[35]. From Figure 4, 5 and 6 (for 30% summary length) it is clear that we are getting high Precision, F-Score w.r.t. MEAD reference summary and high Recall w.r.t Microsoft generated summary.

*Table 5: Summary generated by our algorithm considered as system summary another summary as a model summary.*

| MEASURE | MEAD - 10% | | | MICRO SOFT-10% | | | OPIOSIS-10% | | |
|---|---|---|---|---|---|---|---|---|---|
| **10 % summary** | **R** | **P** | **F** | **R** | **P** | **F** | **R** | **P** | **F** |
| **ROUGE-1** | 0.46 | 0.71 | 0.56 | 0.345 | 0.275 | 0.306 | 0.508 | 0.464 | 0.485 |
| **ROUGE-2** | 0.35 | 0.54 | 0.42 | 0.032 | 0.025 | 0.025 | 0.258 | 0.235 | 0.246 |
| **ROUGE-3** | 0.33 | 0.51 | 0.4 | 0 | 0 | 0 | 0.209 | 0.19 | 0.199 |
| **ROUGE-4** | 0.32 | 0.4.7 | 0.39 | 0 | 0 | 0 | 0.187 | 0.17 | 0.178 |
| **ROUGE-5** | 0.32 | 0.49 | 0.38 | 0 | 0 | 0 | 0.171 | 0.156 | 0.163 |
| **ROUGE-6** | 0.31 | 0.48 | 0.38 | 0 | 0 | 0 | 0.155 | 0.141 | 0.147 |
| **ROUGE-7** | 0.31 | 0.47 | 0.37 | 0 | 0 | 0 | 0.138 | 0.126 | 0.132 |
| **ROUGE-8** | 0.3 | 0.47 | 0.36 | 0 | 0 | 0 | 0.122 | 0.111 | 0.116 |
| **ROUGE-9** | 0.29 | 0.46 | 0.35 | 0 | 0 | 0 | 0.105 | 0.095 | 0.1 |
| **ROUGE-10** | 0.29 | 0.45 | 0.35 | 0 | 0 | 0 | 0.088 | 0.08 | 0.084 |
| **ROUGE-L** | 0.45 | 0.69 | 0.54 | 0.307 | 0.244 | 0.272 | 0.474 | 0.433 | 0.453 |
| **ROUGE-W** | 0.03 | 0.35 | 0.06 | 0.017 | 0.076 | 0.029 | 0.041 | 0.154 | 0.065 |
| **ROUGE-S*** | 0.16 | 0.37 | 0.22 | 0.097 | 0.061 | 0.075 | 0.22 | 0.183 | 0.2 |
| **ROUGE-SU*** | 0.16 | 0.37 | 0.22 | 0.1 | 0.063 | 0.077 | 0.223 | 0.186 | 0.203 |

*Table 6: Summary generated by our algorithm as system summary, another summary as model summary*

| MEASURE | MEAD - 20% | | | OPIOSIS -20% | | |
|---|---|---|---|---|---|---|
| **21 % summary** | **R** | **P** | **F** | **R** | **P** | **F** |
| **ROUGE-1** | 0.399 | 0.619 | 0.485 | 0.508 | 0.521 | 0.515 |
| **ROUGE-2** | 0.251 | 0.39 | 0.306 | 0.267 | 0.273 | 0.27 |
| **ROUGE-3** | 0.222 | 0.345 | 0.27 | 0.207 | 0.212 | 0.21 |
| **ROUGE-4** | 0.214 | 0.334 | 0.261 | 0.181 | 0.185 | 0.183 |
| **ROUGE-5** | 0.211 | 0.329 | 0.257 | 0.166 | 0.17 | 0.168 |
| **ROUGE-6** | 0.207 | 0.323 | 0.253 | 0.151 | 0.155 | 0.153 |
| **ROUGE-7** | 0.204 | 0.318 | 0.248 | 0.137 | 0.14 | 0.138 |
| **ROUGE-8** | 0.2 | 0.313 | 0.244 | 0.122 | 0.125 | 0.123 |
| **ROUGE-9** | 0.196 | 0.308 | 0.24 | 0.107 | 0.11 | 0.108 |
| **ROUGE-10** | 0.193 | 0.302 | 0.235 | 0.092 | 0.094 | 0.093 |
| **ROUGE-L** | 0.391 | 0.607 | 0.475 | 0.485 | 0.496 | 0.49 |
| **ROUGE-W** | 0.027 | 0.287 | 0.05 | 0.04 | 0.167 | 0.065 |
| **ROUGE-S\*** | 0.151 | 0.365 | 0.214 | 0.201 | 0.211 | 0.206 |
| **ROUGE-SU\*** | 0.152 | 0.367 | 0.215 | 0.203 | 0.203 | 0.208 |

*Table 7: Summary generated by our algorithm as system summary, another summary as model summary*

| MEASURE | MEAD - 25 | | | Mead-30 | | | MICRO 25 | | | MICRO 30 | | | OPINIOUS | | | 30% |
|---|---|---|---|---|---|---|---|---|---|---|---|---|---|---|---|---|
| **27 % summary** | R | P | F | R | P | F | R | P | F | R | P | F | R | P | F | |
| ROUGE-1 | 0.49 | 0.69 | 0.57 | 0.42 | 0.70 | 0.53 | 0.64 | 0.57 | 0.60 | 0.55 | 0.61 | 0.58 | 0.50 | 0.54 | 0.52 |
| ROUGE-2 | 0.35 | 0.49 | 0.41 | 0.30 | 0.50 | 0.37 | 0.45 | 0.40 | 0.42 | 0.36 | 0.41 | 0.38 | 0.24 | 0.27 | 0.25 |
| ROUGE-3 | 0.33 | 0.46 | 0.38 | 0.28 | 0.46 | 0.35 | 0.39 | 0.35 | 0.37 | 0.32 | 0.35 | 0.34 | 0.18 | 0.20 | 0.19 |
| ROUGE-4 | 0.32 | 0.45 | 0.37 | 0.27 | 0.45 | 0.34 | 0.37 | 0.34 | 0.35 | 0.30 | 0.34 | 0.32 | 0.15 | 0.17 | 0.16 |
| ROUGE5 | 0.31 | 0.44 | 0.36 | 0.26 | 0.44 | 0.33 | 0.36 | 0.33 | 0.34 | 0.29 | 0.33 | 0.31 | 0.14 | 0.15 | 0.15 |
| ROUGE6 | 0.30 | 0.43 | 0.36 | 0.26 | 0.43 | 0.32 | 0.35 | 0.32 | 0.34 | 0.28 | 0.32 | 0.30 | 0.13 | 0.14 | 0.13 |
| ROUGE7 | 0.30 | 0.42 | 0.35 | 0.25 | 0.42 | 0.32 | 0.34 | 0.31 | 0.33 | 0.28 | 0.31 | 0.29 | 0.11 | 0.12 | 0.12 |
| ROUGE-8 | 0.29 | 0.41 | 0.34 | 0.25 | 0.41 | 0.31 | 0.34 | 0.30 | 0.32 | 0.27 | 0.30 | 0.29 | 0.10 | 0.11 | 0.10 |
| ROUGE-9 | 0.28 | 0.41 | 0.33 | 0.24 | 0.41 | 0.30 | 0.33 | 0.30 | 0.31 | 0.26 | 0.30 | 0.28 | 0.09 | 0.10 | 0.09 |
| ROUGE-10 | 0.28 | 0.40 | 0.33 | 0.23 | 0.40 | 0.29 | 0.32 | 0.29 | 0.30 | 0.26 | 0.29 | 0.27 | 0.07 | 0.08 | 0.08 |
| ROUGE-L | 0.47 | 0.67 | 0.56 | 0.41 | 0.69 | 0.51 | 0.60 | 0.54 | 0.57 | 0.51 | 0.57 | 0.54 | 0.47 | 0.51 | 0.49 |
| ROUGE-W | 0.03 | 0.30 | 0.06 | 0.03 | 0.29 | 0.05 | 0.05 | 0.26 | 0.08 | 0.04 | 0.26 | 0.07 | 0.04 | 0.15 | 0.06 |
| ROUGE-S\* | 0.22 | 0.44 | 0.29 | 0.17 | 0.47 | 0.25 | 0.39 | 0.32 | 0.35 | 0.29 | 0.36 | 0.32 | 0.19 | 0.23 | 0.21 |

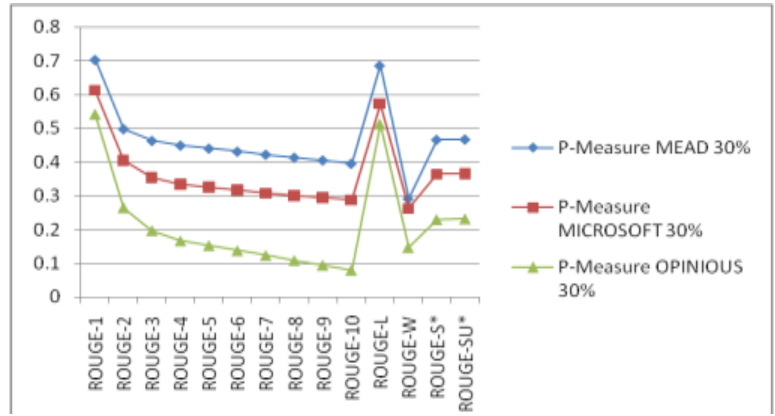

Figure 4: Precision curve

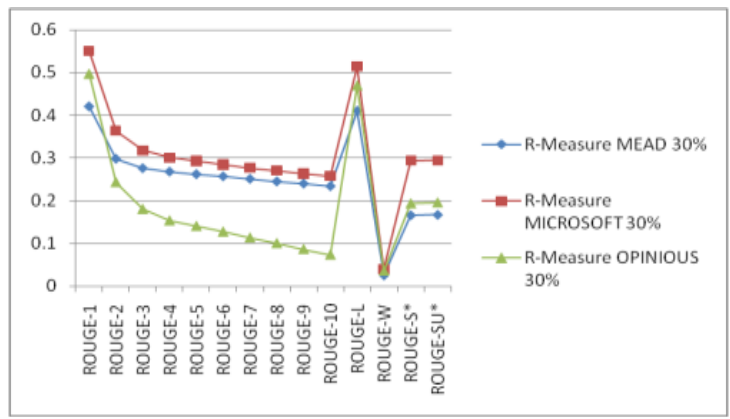

Figure 5: Recall curve

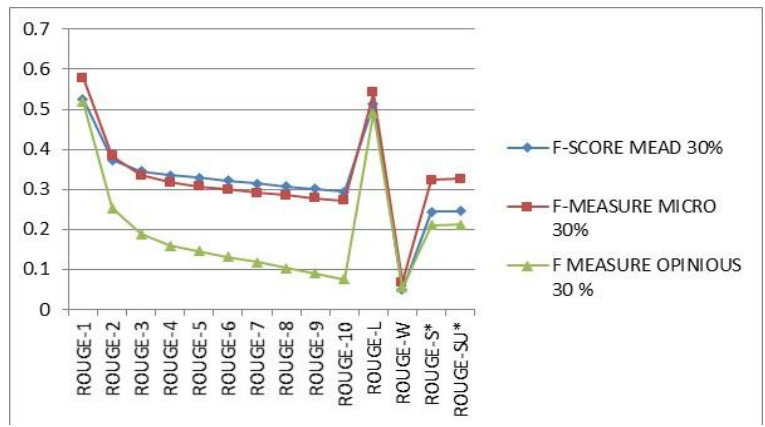

Figure 6: Showing F-Score

## 7.2. Experiment 2

In this experiment we comparing different system generated summary w.r.t. Human summary (or in other words here Model (Gold/ Reference) summary is Human generated summary) and other summaries are system generated summary (i.e. MEAD, MICROSOFT, OPINOSIS, OUR-ALGO are system generated summary). Result are shown in Table 8, and Table 9. Table 8 is representing different ROUGE scores for the summary length of 24%.

*Table 8: Summary generated by different as system summary, human generated summary as model summary.*

| MEASURE USER Summary-24 % | My-Method 24 | | | MEAD - 24 | | | MICRO-24% | | | OPINIOUS -24% | | |
|---|---|---|---|---|---|---|---|---|---|---|---|---|
| | R | P | F | R | P | F | R | P | F | R | P | F |
| ROUGE-1 | 0.45 | 0.49 | 0.47 | 0.53 | 0.38 | 0.44 | 0.07 | 0.59 | 0.12 | 0.47 | 0.50 | 0.48 |
| ROUGE-2 | 0.18 | 0.20 | 0.19 | 0.21 | 0.14 | 0.17 | 0.01 | 0.13 | 0.03 | 0.17 | 0.18 | 0.17 |
| ROUGE-3 | 0.11 | 0.12 | 0.11 | 0.10 | 0.07 | 0.08 | 0.00 | 0.03 | 0.00 | 0.09 | 0.09 | 0.09 |
| ROUGE-4 | 0.07 | 0.08 | 0.08 | 0.06 | 0.04 | 0.05 | 0.00 | 0.00 | 0.00 | 0.06 | 0.06 | 0.06 |
| ROUGE-5 | 0.05 | 0.06 | 0.06 | 0.04 | 0.03 | 0.03 | 0.00 | 0.00 | 0.00 | 0.05 | 0.05 | 0.05 |
| ROUGE-6 | 0.04 | 0.04 | 0.04 | 0.02 | 0.02 | 0.02 | 0.00 | 0.00 | 0.00 | 0.03 | 0.03 | 0.03 |
| ROUGE-7 | 0.03 | 0.03 | 0.03 | 0.01 | 0.01 | 0.01 | 0.00 | 0.00 | 0.00 | 0.02 | 0.02 | 0.02 |
| ROUGE-8 | 0.02 | 0.02 | 0.02 | 0.01 | 0.01 | 0.01 | 0.00 | 0.00 | 0.00 | 0.01 | 0.02 | 0.01 |
| ROUGE-9 | 0.01 | 0.01 | 0.01 | 0.00 | 0.00 | 0.00 | 0.00 | 0.00 | 0.00 | 0.01 | 0.01 | 0.01 |
| ROUGE-10 | 0.01 | 0.01 | 0.01 | 0.00 | 0.00 | 0.00 | 0.00 | 0.00 | 0.00 | 0.01 | 0.01 | 0.01 |
| ROUGE-L | 0.41 | 0.45 | 0.43 | 0.49 | 0.35 | 0.41 | 0.06 | 0.54 | 0.11 | 0.44 | 0.46 | 0.45 |
| ROUGE-W | 0.03 | 0.13 | 0.04 | 0.03 | 0.11 | 0.05 | 0.01 | 0.20 | 0.01 | 0.03 | 0.13 | 0.05 |
| ROUGE-S* | 0.20 | 0.24 | 0.22 | 0.28 | 0.14 | 0.18 | 0.00 | 0.29 | 0.01 | 0.16 | 0.18 | 0.17 |
| ROUGE-SU* | 0.20 | 0.24 | 0.22 | 0.28 | 0.14 | 0.19 | 0.00 | 0.30 | 0.01 | 0.16 | 0.19 | 0.17 |

*Table 9: Summary generated by different system considered as system summary, human generated summary as model summary*

| MEASURE User summary 40 % | My-Method-40% | | | MEAD - 40% | | | MICRO-40% | | | OPINIOUS-40% | | |
|---|---|---|---|---|---|---|---|---|---|---|---|---|
| | R | P | F | R | P | F | R | P | F | R | P | F |
| ROUGE-1 | 0.60 | 0.71 | 0.65 | 0.82 | 0.58 | 0.68 | 0.70 | 0.72 | 0.71 | 0.48 | 0.61 | 0.54 |
| ROUGE-2 | 0.45 | 0.53 | 0.48 | 0.69 | 0.48 | 0.57 | 0.53 | 0.54 | 0.54 | 0.23 | 0.29 | 0.26 |
| ROUGE-3 | 0.40 | 0.48 | 0.44 | 0.64 | 0.45 | 0.53 | 0.48 | 0.49 | 0.48 | 0.15 | 0.19 | 0.17 |
| ROUGE-4 | 0.38 | 0.45 | 0.41 | 0.62 | 0.43 | 0.51 | 0.46 | 0.47 | 0.46 | 0.12 | 0.15 | 0.13 |
| ROUGE-5 | 0.36 | 0.42 | 0.39 | 0.60 | 0.42 | 0.50 | 0.44 | 0.45 | 0.45 | 0.11 | 0.13 | 0.12 |
| ROUGE-6 | 0.34 | 0.40 | 0.37 | 0.59 | 0.41 | 0.48 | 0.42 | 0.44 | 0.43 | 0.09 | 0.12 | 0.10 |
| ROUGE-7 | 0.32 | 0.38 | 0.35 | 0.57 | 0.40 | 0.47 | 0.41 | 0.42 | 0.41 | 0.08 | 0.10 | 0.09 |
| ROUGE-8 | 0.30 | 0.36 | 0.33 | 0.56 | 0.39 | 0.46 | 0.39 | 0.41 | 0.40 | 0.07 | 0.08 | 0.07 |
| ROUGE-9 | 0.28 | 0.34 | 0.31 | 0.54 | 0.38 | 0.45 | 0.38 | 0.39 | 0.38 | 0.06 | 0.07 | 0.06 |
| ROUGE-10 | 0.27 | 0.32 | 0.29 | 0.53 | 0.37 | 0.44 | 0.36 | 0.38 | 0.37 | 0.05 | 0.06 | 0.05 |
| ROUGE-L | 0.58 | 0.69 | 0.63 | 0.82 | 0.57 | 0.67 | 0.69 | 0.72 | 0.70 | 0.47 | 0.59 | 0.52 |
| ROUGE-W | 0.03 | 0.23 | 0.06 | 0.06 | 0.22 | 0.09 | 0.04 | 0.25 | 0.07 | 0.02 | 0.16 | 0.04 |
| ROUGE-S* | 0.344 | 0.48 | 0.4 | 0.68 | 0.337 | 0.5 | 0.49 | 0.523 | 0.51 | 0.184 | 0.294 | 0.227 |
| ROUGE-SU* | 0.345 | 0.48 | 0.4 | 0.69 | 0.337 | 0.5 | 0.49 | 0.524 | 0.51 | 0.185 | 0.295 | 0.228 |

From Table 8 we can say that,

1. We are getting high F-Score comparison to MEAD, MICROSOFT system and OPINOSIS system. Except ROUGE-W in MEAD's ROUGE-1 and OPINOSIS's ROUGE-W. F-score is represent in Figure7.

2. We are getting high PRECISION compare to MEAD and OPINOSIS but, Microsoft system leading in ROUGE-1, ROUGE-L, ROUGE-W, ROUGE-S*, ROUGE-SU* only.

3. We are getting High RECALL comparison to MEAD in ROUGE-3 to ROUGE-10 and higher Recall comparison to OPINOSIS and MICROSOFT in all measures except OPINOSIS getting ROUGE-1 higher than OUR-System.

In Figure 7 we are representing comparison of different system generated summary (24 % length) using F-Measure, and Figure 8 (comparison of 40% length summary ) and we representing here only F-Score. From Table 9 we can say that,

4. MEAD system and MICROSOFT performing better in term of RECALL, but our system is performing better compare to OPINOSIS.

5. OUR Method getting Higher PRECISION compare to OPINSIS's all ROUGE score (P) and higher PRECISION achieved compare to MEAD except ROUGE-6 to ROUGE-10.

6. We are getting low F-Score compare to MEAD and MICROSOFT system but higher w.r.t OPINOSIS.

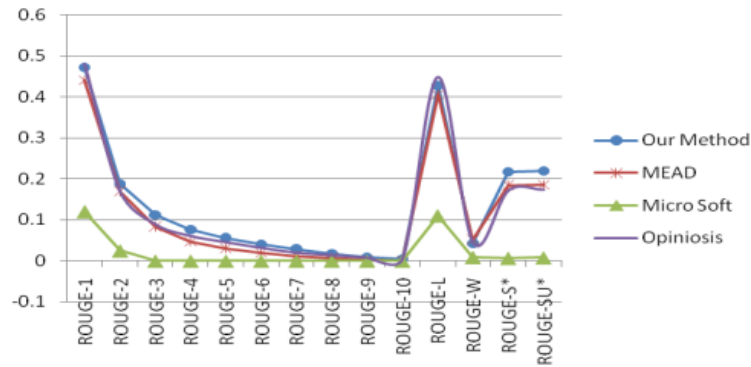

Figure 7: F-score 24 % summary

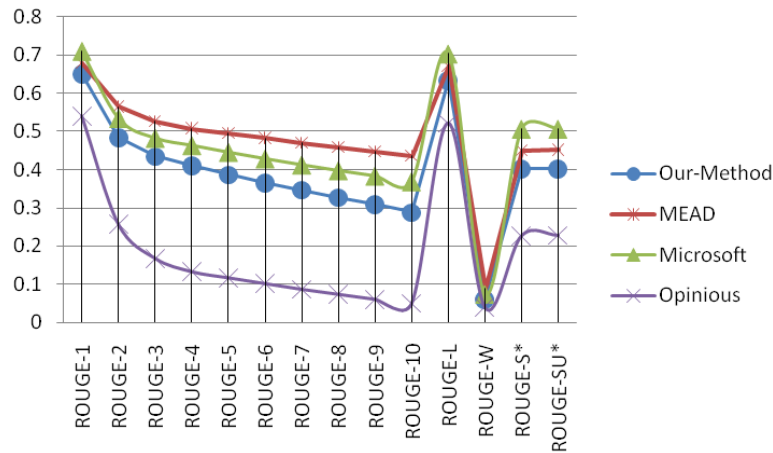

Figure 8: F-Score 40% summary

### 7.3. Experiment 3

In this Experiment we are showing the significance of Sentiment feature, the purpose of this experiment to show is really sentiment score performing a significant role in salient sentence extraction?. To generate a good quality summary of some words (like 100 words) is a tedious task. In our experiment we are taken five different features. We tried all combination of all five features, and using this combination we trying to prove this feature is playing a significant role in summarization.

If number of features are n then, the total number of combination possible=$2^n$-1, so in our experiment we are trying all 31 combination (calling 31 approaches). Here we generate summary using single stand alone feature based summary, and summary in which sentiment score is playing a role. Here first we generate approximate 100 words summary, to evaluate this summary we took three human generated summary as Gold/ Reference summary. Motivated by DUC -2002 task, we are evaluating only first 100 words of the summary.

Let:1- stands for TF-IDF feature ; 2-stands for Aggregate similarity score; 3-stands for Position based feature;4- stands for Centroid based feature; 5-stands for Sentiment based score; 1+2+3 feature showing collective score of TF-IDF, Aggregate similarity, Position based features.

*Table 10 : Different ROUGE score for summary generated using different Approaches.*

| Approach | 1 | 2 | 3 | 4 | 5 | 9 | 12 | 14 | 15 | 6 | 18 | 10 | 20 |
|---|---|---|---|---|---|---|---|---|---|---|---|---|---|
| Measures | 1 | 2 | 3 | 4 | 5 | 1+5 | 2+5 | 3+5 | 4+5 | 1+2 | 1+2+5 | 2+3 | 2+3+5 |
| ROUGE-1 | 0.563 | 0.279 | 0.622 | 0.265 | 0.566 | 0.576 | 0.569 | 0.585 | 0.563 | 0.56 | 0.582 | 0.625 | 0.582 |
| ROUGE-2 | 0.459 | 0.077 | 0.5 | 0.056 | 0.427 | 0.433 | 0.429 | 0.433 | 0.447 | 0.457 | 0.433 | 0.504 | 0.433 |
| ROOUGE-L | 0.54 | 0.242 | 0.592 | 0.216 | 0.537 | 0.533 | 0.54 | 0.54 | 0.53 | 0.537 | 0.533 | 0.595 | 0.536 |
| ROUGE-W | 0.319 | 0.105 | 0.342 | 0.092 | 0.315 | 0.314 | 0.316 | 0.316 | 0.314 | 0.318 | 0.313 | 0.343 | 0.315 |
| ROUGE-S* | 0.324 | 0.065 | 0.403 | 0.059 | 0.34 | 0.348 | 0.34 | 0.361 | 0.329 | 0.321 | 0.357 | 0.409 | 0.358 |
| ROUGE-SU* | 0.329 | 0.069 | 0.407 | 0.062 | 0.344 | 0.352 | 0.348 | 0.365 | 0.334 | 0.326 | 0.362 | 0.413 | 0.362 |

| Approach | 13 | 21 | 19 | 23 | 17 | 24 | 22 | 26 | 16 | 27 |
|---|---|---|---|---|---|---|---|---|---|---|
| Measures | 3+4 | 3+4+5 | 2+3+4 | 2+3+4+5 | 1+2+4 | 1+2+4+5 | 1+2+3+4 | 1+2+3+4+5 | 1+2+3 | 1+2+3+5 |
| ROUGE-1 | 0.504 | 0.563 | 0.508 | 0.563 | 0.284 | 0.566 | 0.504 | 0.563 | 0.553 | 0.595 |
| ROUGE-2 | 0.35 | 0.446 | 0.355 | 0.447 | 0.062 | 0.449 | 0.35 | 0.446 | 0.43 | 0.44 |
| ROOUGE-L | 0.454 | 0.523 | 0.459 | 0.524 | 0.232 | 0.527 | 0.454 | 0.523 | 0.527 | 0.545 |
| ROUGE-W | 0.254 | 0.312 | 0.258 | 0.312 | 0.098 | 0.313 | 0.254 | 0.312 | 0.313 | 0.318 |
| ROUGE-S* | 0.201 | 0.329 | 0.204 | 0.328 | 0.065 | 0.331 | 0.201 | 0.329 | 0.33 | 0.367 |
| ROUGE-SU* | 0.207 | 0.334 | 0.21 | 0.333 | 0.069 | 0.336 | 0.207 | 0.334 | 0.334 | 0.372 |

Here we are presenting 23 different features combination to find a summary. By seeing Table 10, we can see that Position based feature (approach 3- highlighted in green) is performing best among all, but this is due to, that in all three human reference summary (out of 5 summary 3 are extractive type summary- extractive type summary is available at address 36 ), which are used for evaluation contains top sentences 1, 3, and 4 which is almost 100 words, and the model which we are using for score position giving higher preference for leading sentences, but as we know position based score can't perform well in all cases like in scientific article so we need some more features. From Table 10 this is clear when we are taking Sentiment feature (id 5), along with other features we are getting improved summary. More Rouge score means more accurate summary. The conclusion of this experiment is that,

➢ Out of 11 approaches (when sentiment feature added), 9 times we are getting improved summary by adding Sentiment feature. for example, If we take collective features (1+2+3+4), and by adding sentiment feature (1+2+3+4+5) we are getting improved results (highlighted in red color).

➢ Here Position based feature performing best among all approaches, the reason is given above and we can't depend only on Position based feature so we need more features.

➢ In Approach 10 (2+3 i.e. Aggregate and Position), when we add Sentiment score (done in approach 20 i.e. 2+3+5, highlighted in Blue color), the performance is reduced, this be due to Position based score not preferred i.e. Position based feature is not dominating here as in Approach 3.

Results obtain from three human summary as Reference summary, and summary obtain from different 31 approaches consider as system summary, along with document are available at [36] address. To remove biasness and evaluate first 100 world summary we use -l 100, and to evaluate ROUGE-W we have taken W=1.2.

# 8.    Conclusion and future work

In this work we are taken dataset designed by us. We presented a hybrid text document summarization algorithm based on linear combination of different statistical measures and semantic measures. In our hybrid approach we taken statistical measures like sentence position, centroid, TF-IDF as well as semantic approach (doing sentiment analysis) that is based on word level analysis. Sentiment score of a sentence is given as the sum of sentiment score of every entities present in a sentence. We are getting three polarity for any entity like Neutral, Negative and Positive. If entity sentiment is negative then we multiplying every score by -1 to treat it as positive score , the reason for doing this we wants to select a sentence in which strong sentiment is present it may be either negative or positive and both have same importance for us. The significance/contribution of sentiment score is presented in section 6.3/ experiment-3. We are not using any learning or brute force approach to deciding how much importance to given different measures or in other words we giving equal importance for each and every feature. To calculate the score for a sentence we just add all the scores for every sentence and pick up a sentence based on highest score. In next step, we select next sentence based on next highest score and add it to the summary if the similarity between summary and sentence is lower that threshold to maintain redundancy and coverage. We stop our algorithm when the desired length summary is found.

To generate several summaries of different length we used methods like MEAD, Microsoft, OPINOSIS and HUMAN and for evaluation different ROUGE score is used. We done two experiment in first experiment, we take our summary (generate from ALGORITHM described in section 5 ) as system summary  and all other as model summary and it have showed that we getting high precision almost every time,  that denotes we covered most relevant results.  In the second experiment we compare different system generated summary (MEAD, Microsoft, OPINOSIS, OUR ALGORITHM) to Model summary (HUMAN GENERATED). In this we find that our explained algorithm performed well for 24% generated summary for almost every time but, in 40 % MEAD system generate summary leading in some way but here also we getting higher RECALL compare to MEAD. In third experiment, section 6.3 we have shown that when we are adding sentiment score as a feature we are getting improved results compare to without sentiment score. Third experiment is showing that sentiment score has contribution in extraction of more appropriate sentences

Limitation of our work that, in Step 2 i.e. salient sentence extraction we are initializing two parameter L-desired summary length and θ Similarity Threshold,  we need to set θ very less .04. If we set θ as .5 or .6 the sentences which came in summary only depends on the output of Step 2, which are not following one important property of summary that is "Coverage". Since we are using long stop word list and wants to follow coverage property we need to choose θ with precaution but we can put high θ if not using any Stop-Word list. In future (1) we can use soft computing technique to learn different weights for different feature scores, (2) we can extend our approach to Multi Document summarization, (3) we can extend this experiment for benchmark dataset.

# 9.    Acknowledgement

Thanks to UGC for funding me, to Iskandar Keskes (Miracl laboratory, ANLP-Research Group, Sfax-Tunisia) to give me description about how to evaluate summary. Thanks to Mr. Prem Ayadav, Mr. Sarad, Ms. Payal Biswas to generate extractive type summary from given document , and Ashish Kumar (All from SC & SS, IR-LAB 01, JNU, Delhi) to help me at several stages.